\documentclass[aps,prl,reprint,preprintnumbers,superscriptaddress,secnumarabic,amsmath,amssymb,floatfix,nofootinbib,final]{revtex4-2}
\pdfoutput=1
\usepackage{abv-prl}

\usepackage{comment}
\usepackage{placeins}
\usepackage{color}
\usepackage[normalem]{ulem}

\usepackage{mathtools}
\usepackage{upgreek}
\usepackage{empheq}
\usepackage[many]{tcolorbox}

\usepackage[compat=1.1.0]{tikz-feynman}

\renewcommand{\arraystretch}{1.6}
\theoremstyle{definition}

\makeatletter
\newtcolorbox{FJBlock}{
  enhanced,
  breakable,
  parbox=false,
  code={\let\tcb@set@parbox@indent\@empty},
  before upper={\@minipagefalse},
  colback=rust!6!white,
  coltext=black,
  frame hidden,
  boxrule=0pt,
  sharp corners,
  boxsep=0pt,
  left=0pt, right=0pt,
  top=0pt, bottom=0pt,
  beforeafter skip balanced=0pt
}
\makeatother

\newcommand{\vac}{\mathrm{vac}}

\newcommand{\bdy}{\mathrm{bdy}}
\newcommand{\sing}{\mathrm{sing}}
\newcommand{\hor}{\mathrm{hor}}

\newcommand{\src}{\mathrm{src}}
\newcommand{\res}{\mathrm{res}}

\begin{document}
\preprint{{CERN-TH-2026-262}}
\title{%
Resurgence of Black Holes in Thermal OPE
}
\author{Alba Grassi}

\affiliation{Section de Math\'ematiques, Universit\'e de Gen\`eve,
1211 Gen\`eve 4, Switzerland}
\affiliation{Theoretical Physics Department, CERN,
1211 Geneva 23, Switzerland}
\author{Hewei Frederic Jia}
\affiliation{Institute for Advanced Study, Tsinghua University,
Beijing 100084, China}

\begin{abstract}
The momentum-space thermal OPE generically gives an asymptotic high-energy expansion of thermal two-point functions. We study the resurgent structure of this expansion in holographic CFTs at infinite and finite volume, whose dual bulk geometries are planar and spherical AdS black holes, respectively.  The perturbative expansion is fixed to all orders by the stress-tensor sector of the thermal OPE and can be computed from a near-boundary analysis of the bulk wave equation.  Non-perturbative effects, on the other hand, encode information about the bulk geometry away from the boundary. Using exact WKB analysis, we identify the locations of the associated Borel singularities as WKB actions between the AdS boundary and distinguished points in the complex radial plane, which include the black hole singularity in certain regimes. Furthermore, we identify a complex-frequency chamber in the timelike momentum regime where Borel resummation of the high-energy expansion reconstructs the full thermal two-point function. We further derive Stokes jumps that determine the non-perturbative contributions required upon crossing the boundary of this chamber.
\\\newline

\end{abstract}

\maketitle

\section{Introduction and Summary}

Thermal two-point functions describe the response of quantum field theories at finite temperature. At high energies, their momentum-space expansion is controlled by the thermal operator product expansion (OPE) \cite{Caron-Huot:2009ypo,Katz:2014rla,Iliesiu:2018fao}.
Although the position-space OPE converges in CFTs, its momentum-space counterpart is generally expected to be asymptotic \cite{ShifmanVainshteinZakharov:1979Foundations,Shifman:1998Snapshots,Manenti:2019wxs}.

This raises several natural questions. What controls the large-order behavior of the high-energy expansion? When its coefficients grow factorially, what physical information is encoded in the associated Borel singularities, and under what condition does Borel resummation reconstruct the full thermal correlator?

The AdS/CFT correspondence provides a controlled setting in which to address these questions. In this work we focus on thermal two-point functions in holographic \CFT{d} on spatial
$\mathbb{R}^{d-1}$ and $S^{d-1}$, dual to neutral \AdS{d+1}
black branes and black holes, respectively.\footnote{We work in
the black hole phase for \CFT{d} on spatial $S^{d-1}$.}
Thermal correlators in these theories are computed by studying
perturbations of the dual black hole backgrounds \cite{Son:2002sd}.

We consider the high-energy limit in which frequency and spatial
momentum become large at a fixed ratio. The resulting perturbative
expansion is determined to all orders by the stress-tensor sector
of the thermal OPE and can be computed holographically from a
near-boundary analysis of the bulk wave equation
\cite{Fitzpatrick:2019zqz,AfkhamiJeddi:2025bh}.
Determining the full correlator, however, requires additional non-perturbative information about the bulk geometry away
from the boundary. We extract this information from the perturbative high-energy expansion using resurgence and exact WKB analysis.

We classify the principal-sheet Borel singularities through saddle trajectories in the Stokes graph reaching the AdS boundary in the complexified radial plane. The associated WKB actions determine their locations. In particular, we identify the kinematic regimes where saddle trajectories reaching the black hole singularity contribute to the Borel structure.

We also identify a complex-frequency reconstruction chamber within the timelike momentum regime, where Borel resummation of the thermal OPE reproduces the full retarded correlator. We derive an explicit Stokes jump that fixes the additional non-perturbative factor required when crossing the boundary of this chamber.

  \section{Thermal correlator and black hole}
We study the thermal retarded two-point function of scalar primary operators of dimension $\Delta$ in \CFT{d}\footnote{Likewise the advanced correlator is defined by
$G_{\mathrm A}(t,x)=-i\,\theta(-t)
\langle[\mathcal O(t,x),\mathcal O(0,0)]\rangle_\beta$.}
\begin{equation}
G_{\mathrm R}(t,x)
=
i\,\theta(t)
\left\langle
[\mathcal O(t,x),\mathcal O(0,0)]
\right\rangle_\beta.
\end{equation}
We denote its temporal Fourier transform and spatial Fourier or spherical-harmonic transform by $G_{\mathrm R}(\omega,k)$ or $G_{\mathrm R}(\omega,\ell)$. We consider the high-energy limit at fixed momentum-to-frequency ratio
\begin{equation}
|\beta\omega|\longrightarrow\infty,
\qquad
\chi^2=
\begin{cases}
\dfrac{k^2}{\omega^2},
&\mathbb R^{d-1},\\[6pt]
\dfrac{\ell(\ell+d-2)}{R^2\omega^2},
&S_R^{d-1},
\end{cases}
\quad\text{fixed}.
\label{eq:high-energy}
\end{equation}
The perturbative corrections to the vacuum  two-point function are controlled by thermal OPE data for generic \CFT{d}~\cite{Iliesiu:2018fao}. For holographic \CFT{d}, only the multi-stress-tensor sector contribute to the high-energy expansion. For example, for spatial $\mathbb{R}^{d-1}$, its explicit form reads 
\begin{equation}
\begin{aligned}
&\frac{G_{\mathrm R}^{\mathrm{pert}}(\omega,k)}
     {G_{\mathrm R}^{\vac}(\omega,k)}
\\
&=1+
\sum_{n\ge1}
\frac{2^{nd}}
{\beta^{nd}(\omega^2-k^2)^{nd/2}}
\sum_{\substack{0\le J\le2n\\J\ \mathrm{even}}}
\frac{a_{T^n,J}}{a_{\mathbf1}}
\\
&\times
e^{i\pi(nd-J)/2}
\frac{
\Gamma(\Delta)\,
\Gamma\!\left(\frac{(n+1)d+J}{2}-\Delta\right)}
{
\Gamma(d/2-\Delta)\,
\Gamma(\Delta-nd/2+J/2)}
\\
&\times
C_J^{((d-2)/2)}
\!\left(\frac{\omega}{\sqrt{\omega^2-k^2}}\right).
\end{aligned}
\end{equation}
where $a_{T^n,J}$ denote thermal OPE coefficients of
$n$-stress-tensor operators of dimension $nd$ and spin $J$, in the
convention of \cite{JiaRangamani:2026Exact}, and $a_{\mathbf1}$
is the identity coefficient. The convention for $G_R^{\vac}$ is given in \eqref{eq:short-vacuum-normalization}.

We will focus on holographic \CFT{d} with even $d$.\footnote{We restrict to even $d$, since for odd $d$ the singularity structure of the canonical Schr\"odinger equation falls outside the standard assumptions of exact WKB analysis. Moreover, the near-boundary expansion is more involved for odd $d$~\cite{Fitzpatrick:2019zqz}. We also take $\nu\notin\mathbb Z$ to avoid resonant boundary exponents and logarithmic terms in the local Frobenius solutions. For $\nu\in\mathbb N$ many simplifications occur, see \cite{JiaRangamani:2024Blocks,Jia:2025jbi,JiaRangamani:2026Exact}.} The thermal state in holographic \CFT{d} is dual to an \AdS{d+1} black brane/hole 
\begin{equation}
\begin{aligned}
ds^2
&=\frac{r_+^2}{z}
\left[-f(z)\,dt^2
+d\Sigma_{K,d-1}^2\right]
+\frac{\,dz^2}{4z^2f(z)},\\
f(z)&=1+a_Kz-(1+a_K)z^{d/2},\\
z&=\frac{r_+^2}{r^2},
\qquad
a_K=\frac{K}{r_+^2}.
\end{aligned}\label{eq:metric}
\end{equation}
Here $K=0$ and $K=1$ correspond to the black brane and black hole, respectively, with $d\Sigma_{K,d-1}^2$ denoting the metric on $\mathbb{R}^{d-1}$ and the unit sphere $S^{d-1}$, respectively.
The horizon $r_+$ is related to the temperature by  $\beta=4\pi/
[r_+(d+(d-2)a_K)]$. 
In this setting, a scalar operator of dimension $\Delta$ on the boundary CFT, is dual to a scalar field in the bulk with mass $m^2=\Delta(\Delta-d)$; we also denote $\nu = \Delta-d/2$.
We write the bulk wave equation in a Schr\"odinger form
\begin{equation}
 \left[\frac{\dd^2}{\dd z^2}-\eta^2Q_0(z)-Q_2(z)\right]\psi(z)=0,
 \label{eq:short-radial-WKB}
\end{equation}
with WKB parameter $\eta = \beta\omega (d+(d-2)a_K)/4\pi$, and
\begin{equation}
  Q_0(z)=\frac{\chi^2f(z)-1}{4zf(z)^2}.
  \label{eq:short-leading-potential}
\end{equation}
The subleading potential $Q_2$ depends on $\nu$, see \eqref{eq:subleading}.
The ingoing and outgoing solutions of \eqref{eq:short-radial-WKB} are specified at the horizon by
\begin{equation}
\begin{aligned}
\psi_{\mathrm{in}}(z)&\sim(1-z)^{1/2-i\beta\omega/(4\pi)},\\
\psi_{\mathrm{out}}(z)&\sim(1-z)^{1/2+i\beta\omega/(4\pi)},
\qquad z\to1.
\end{aligned}
\label{eq:short-horizon-Frobenius}
\end{equation}
At the AdS boundary,  the source and response  solutions  as $z\to0$ are
\begin{equation}
\begin{aligned}
\psi_{\src}(z)&\sim z^{(1-\nu)/2},\quad
\psi_{\res}(z)&\sim z^{(1+\nu)/2}~.
\end{aligned}
\label{eq:short-boundary-asymptotics}
\end{equation}
Their connection to the horizon solutions reads
\begin{equation}
\psi_{\mathrm{in/out}}
=C_{\mathrm{in/out}}^{\src}\psi_{\src}
+C_{\mathrm{in/out}}^{\res}\psi_{\res}.
\label{eq:short-horizon-boundary-connection}
\end{equation}
The holographic prescription for the full retarded and advanced two-point functions is then \cite{Son:2002sd}
\begin{equation}
\begin{aligned}
G_{\mathrm R}
&=\mathcal N_\Delta
\frac{C_{\mathrm{in}}^{\mathrm{res}}}
     {C_{\mathrm{in}}^{\mathrm{src}}}
=\mathcal N_\Delta
\frac{W[\psi_{\src},\psi_{\mathrm{in}}]}
     {W[\psi_{\mathrm{in}},\psi_{\res}]},
\\
G_{\mathrm A}
&=\mathcal N_\Delta
\frac{C_{\mathrm{out}}^{\mathrm{res}}}
     {C_{\mathrm{out}}^{\mathrm{src}}}
=\mathcal N_\Delta
\frac{W[\psi_{\src},\psi_{\mathrm{out}}]}
     {W[\psi_{\mathrm{out}},\psi_{\res}]},
\end{aligned}
\label{short-GR-holo-prescription}
\end{equation}
where $W[u,v]=u\,v'-u'\,v$ and $\mathcal N_\Delta$ fixes the overall normalization of the operator.
To extract holographically their perturbative expansion, it suffices to perform a near-boundary analysis of the bulk wave equation \cite{Fitzpatrick:2019zqz,AfkhamiJeddi:2025bh}, without looking at the behavior close to the horizon. In fact both $G_{\mathrm R/A}^{\mathrm{pert}}/G_{R/A}^{\vac}$ have the same perturbative expansion.

The high-energy limit \eqref{eq:high-energy} corresponds to the WKB limit of \eqref{eq:short-radial-WKB}. Aspects of this limit have been studied in
\cite{FestucciaLiu:2009BohrSommerfeld,Jia:2025jbi,JiaRangamani:2026Exact,Araya:2026shz,Jia:2026pmv,David:2026mlz,Arnaudo:2026der,BarratEtAl:2026MomentumBootstrap,ArnaudoIossaKarlssonWithers:2026OPEQNM,AnicetoArnaudoRatcliffeSpalinski:2026Analytic}, mostly for black branes in \AdS{5}.

The WKB curve and differential associated to \eqref{eq:short-radial-WKB} are
\begin{equation}
   \Sigma_{\mathrm{WKB}}:\qquad
 y^2=Q_0(z),
 \qquad \lambda_0=y\,\dd z.
 \label{eq:short-WKB-curve}
\end{equation}
Collision of bulk turning points occurs at the zeros of the discriminant of the WKB curve. For the black hole, the smallest positive critical value for which the discriminant vanishes is
\begin{equation}
   \chi^{\min}_{\mathrm  c}=\left[
 1+\frac{d-2}{d}a_K
 \left(\frac{2a_K}{d(1+a_K)}\right)^{2/(d-2)}
 \right]^{-1/2} \in (0,1).
 \label{eq:short-critical-chi}
\end{equation}
We denote the largest positive real critical value by $\chi^{\max}_{\mathrm  c}$, see \eqref{eq:chi_max}. For the black brane, both values coincide at the lightlike value, $\chi^{\min}_{\mathrm  c} = \chi^{\max}_{\mathrm  c}=1$.
In this work, for concreteness, we focus on real $\chi$, which suffices to illustrate the main features of the resurgent structure. However, the same methods also apply to complex $\chi$, where we expect a similar resurgent structure.

\section{Borel plane singularities}
We denote the normalized logarithm of the perturbative expansion as 
\begin{equation}
   H(\eta,\chi)=\log\left[
 \frac{G_{\mathrm  R}^{\mathrm  pert}}
      {G_{\mathrm  R}^{\vac}}\right].
 \label{eq:short-normalized-logarithm}
\end{equation}
It can be efficiently computed by extending the method of \cite{AfkhamiJeddi:2025bh} to our setting \cite{WIP}.
The expansion coefficients exhibit factorial growth. For the black brane we have
\begin{equation}
 \begin{aligned}
 H_{\mathrm  bb}(\eta,\chi)
 &=\sum_{n\geq1}H_n^{\mathrm  bb}(\chi)\,\eta^{-dn},
 \\ &\quad  \text{with} \quad H_n^{\mathrm  bb}\sim  \Gamma\left(dn\right) \quad n\gg 1
 \end{aligned}
 \label{eq:short-logarithmic-series-bb}
\end{equation}
while for the black hole we have
\begin{equation}
 \begin{aligned}
 H_{\mathrm  bh}(\eta,\chi)
 &=\sum_{n\geq1}H_n^{\mathrm  bh}(\chi)\,\eta^{-2n},
 \\ &\quad  \text{with} \quad H_n^{\mathrm  bh}\sim  \Gamma(2n) \quad n\gg 1.
 \end{aligned}
 \label{eq:short-logarithmic-series-bh}
\end{equation}
The $nd$-power expansion in the planar case reflects the multi-stress-tensor sectors. For the spherical case, the more general even-power expansion results from curvature corrections and exchanges involving derivatives of the stress tensors \cite{Fitzpatrick:2019zqz}. The factorial growth of the coefficients in
\eqref{eq:short-logarithmic-series-bb} and
\eqref{eq:short-logarithmic-series-bh} motivates the Borel transforms with the Borel variable $s$ conjugate to $\eta$
\begin{equation}
 \begin{gathered}
 \widehat H_{\mathrm  bh}(s,\chi)
 =\sum_{k\geq1}\frac{H_k^{\mathrm  bh}(\chi)}{\Gamma(2k)}s^{2k-1},
 \\
 \widehat H_{\mathrm  bb}(s,\chi)
 =\sum_{k\geq1}\frac{H_k^{\mathrm  bb}(\chi)}{\Gamma(dk)}
 s^{dk-1}.
 \label{eq:short-borel-transforms}
 \end{gathered}
\end{equation}
The Borel singularities in $s$ will be generally complex periods of the classical WKB curve \cite{BalianParisiVoros:1978}, and are therefore associated with Stokes geometry at complex frequency. At phase $\vartheta=\arg \omega = \arg \eta$, the Stokes trajectories emanating from a Stokes source \footnote{We use ``Stokes source'' for zeros of arbitrary order and simple poles of the classical quadratic differential $Q_0 dz^2$.
At non-zero momentum and in $d=6$, the black hole singularity acts as a Stokes source despite being neither a zero nor a pole of the classical differential \cite{Koike:2000ghost}.} $a$  are defined by
\begin{equation}
\operatorname{Im}\!\left[
\re^{\ri\vartheta}\int_a^z\lambda_0
\right]=0.
\end{equation} 
In our problem, the AdS boundary is a Stokes source from which one Stokes trajectory emanates. Other Stokes sources include  bulk turning points in the complexified radial plane and the black hole singularity. At generic phases $\vartheta$, Stokes trajectories flow to (complex) horizons. There exist critical phases $\vartheta_*$ where the Stokes trajectory emanating from the AdS boundary flows into another Stokes source, either some bulk turning point or the black hole singularity. Such a Stokes trajectory is called a saddle trajectory \cite{Voros:1983quartic,Voros:1981bkw}; see~\cite{IwakiNakanishi:2014wkb} for a systematic treatment. It can be shown using exact WKB analysis  that the Borel singularities of \eqref{eq:short-borel-transforms} precisely arise from such saddle trajectories. Let $l_* (\bdy,a)$  be the saddle trajectory at critical phase $\vartheta_*$ between the AdS boundary and another Stokes source $a$. The associated Borel singularity is
\begin{equation}
  s_{*,\pm} = \pm 2 \int_{l_*(\bdy,a)} \lambda_0.
  \label{eq:short-saddle-action}
\end{equation}
In particular, the Borel singularity has phase $\arg s_{*,+} = -\vartheta_*$. We focus on two regimes where the principal-sheet Borel singularities admit a common geometric description for black brane and black hole:
\begin{equation}
\begin{aligned}
\text{Regime I:}\qquad &0\leq\chi<\chi^{\min}_{\mathrm c},\\
\text{Regime II:}\qquad &\chi>\chi^{\max}_{\mathrm c}.
\end{aligned}\label{eq:regimes_borel}
\end{equation}
The Stokes geometry shares the same topology in these two regimes for black brane and black hole. Other regimes in  the black hole case can be systematically analyzed using the same method \cite{WIP}. For concreteness, we present here the explicit form of the saddle trajectories in the black brane case. The black hole example is provided in Sec.~\ref{app:short-spherical-examples} of the Supplementary Material.

In Regime I, for the black brane, the only saddle trajectories are the straight segments \(l_*^{(j)}({\bdy},t_j)=t_j[0,1]\) joining the AdS boundary to the bulk turning points \(t_j\) at critical phases $\vartheta^{(j)}_*(t)$:
\begin{equation}
\begin{aligned}
  \vartheta^{(j)}_*(t)&=
\frac{\pi}{2}-\frac{(2j+1)\pi}{d}
\pmod{\pi},\\
t_j&=
(\chi^{-2}-1)^{2/d}
\re^{2\pi i(2j+1)/d}.
\end{aligned}
\label{eq:regime-i-borel-sing}
\end{equation} 
Here the index range is $j=0, \dots, d/2-1$. The associated  Borel singularities are
\begin{equation}
\begin{aligned}
s_{*,\pm}^{(j)}(\mathrm{bdy},t_j)
={\pm}&\re^{-i\vartheta_*^{(j)}(t)}
\frac{\sqrt{\pi}\,\Gamma(1+1/d)}
{\Gamma(3/2+1/d)}
|1-\chi^2|^{1/2+1/d}
\\
&\times{}_2F_1\!\left(
\frac1d,\frac12+\frac1d;
\frac32+\frac1d;
1-\chi^2
\right).
\end{aligned}
\label{eq:short-brane-regime-i-borel-singularities}
\end{equation}
In the limit $\chi\to0$, the turning points approach the black hole singularity, and the corresponding Borel singularities reduce to
\begin{equation}
\label{eq:short-brane-zero-momentum-borel-singularities}
s_{*,\pm}^{(j)}(\bdy,\sing)
=\pm
\frac{2\pi\,\re^{-\ri\vartheta_*^{(j)}(t)}}
{d\sin(\pi/d)}.
\end{equation}
These may be viewed as a momentum-space version of the bouncing singularity \cite{FidkowskiHubenyKlebanShenker:2003,FestucciaLiu:2005pn,Ceplak:2024bja,Dodelson:2025jff,Parisini:2023nbd, Giombi:2026kdz}. 

In Regime II, there exist both saddle trajectories $l_*^{(j)}(\bdy,t_j)=t_j[0,1]$ joining the AdS boundary to a turning point, and additional ones joining the AdS boundary with the black hole singularity. For the boundary-to-turning-point saddle trajectories, the critical phases and endpoints are
\begin{equation}
\begin{aligned}
\vartheta^{(j)}_*(t) &=-\frac{2\pi j}{d}\pmod{\pi},\\
t_j&=(1-\chi^{-2})^{2/d}\re^{4\pi \ri j/d}.
\end{aligned}\label{eq:crit2}
\end{equation}
The associated Borel singularities are again given by \eqref{eq:short-brane-regime-i-borel-singularities}, with the critical phases \eqref{eq:crit2}. The saddle trajectories associated with the black hole singularity are
\begin{equation}
    l_*^{(j)}(\bdy,\sing)
=\re^{2\pi \ri(2j+1)/d}\mathbb R_+
\end{equation}
at phase
\begin{equation}
\begin{aligned}
\vartheta^{(j)}_*(\sing) &=-\frac{(2j+1)\pi}{d}\pmod{\pi}.
\end{aligned}
\end{equation}
Their associated Borel singularities are
\begin{equation}
\begin{aligned}
s_{*,\pm}^{(j)}&(\bdy,\sing)
={\pm}
\frac{
2\chi\,\Gamma(1+1/d)\Gamma(1/2-1/d)
}{\sqrt{\pi}}
\\
&\times \re^{-\ri\vartheta^{(j)}_*(\sing)}{}_2F_1\!\left(
-\frac12,\frac12-\frac1d;
\frac12;
\chi^{-2}
\right).
\end{aligned}
\label{eq:short-brane-regime-ii-infinity-borel-singularities}
\end{equation}
The index range is $j=0, \dots, d/2-1$ in both cases.

\begin{figure*}[tbp]
 \centering
 \includegraphics[width=\textwidth]{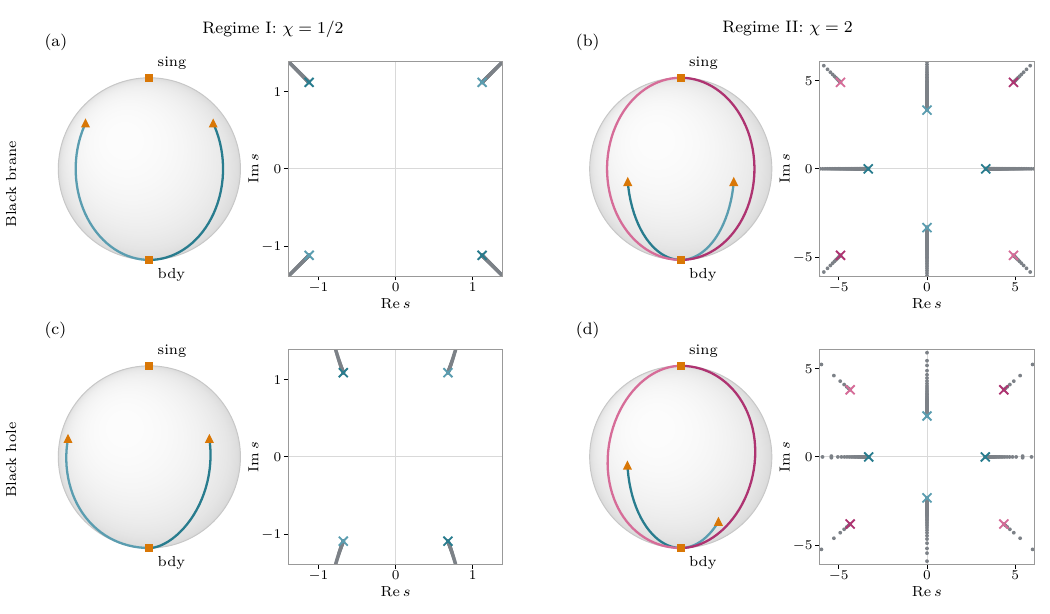}
 \caption{Saddle trajectories on the radial Riemann sphere and their associated Borel singularities for $d=4$, $\nu=13/10$, and $a_K=0$ (top), $a_K=1$ (bottom). Each radial Riemann sphere panel shows saddle trajectories at different critical phases. Matching colors associate each saddle trajectory with its pair of classical-action predictions in the Borel plane (crosses); gray dots are finite Pad\'e poles. The Pad\'e approximants use 256 non-zero coefficients in each panel.}
 \label{fig:borel-saddle-correspondence}
\end{figure*}

Equations \eqref{eq:short-brane-regime-i-borel-singularities} and \eqref{eq:short-brane-regime-ii-infinity-borel-singularities}
locate the leading Borel singularities. Subleading singularities may
occur at integer multiples or combinations of these actions, but are
often obscured in Borel--Pad\'e plots by poles approximating the leading
branch cuts. Examples of Borel planes are given in Fig.~\ref{fig:borel-saddle-correspondence}. A more detailed description of subleading and second-sheet singularities will appear in \cite{WIP}.

\section{Borel reconstruction of thermal correlators}
Given the Borel transforms in \eqref{eq:short-borel-transforms}, we define
the corresponding Borel resummation as \begin{equation}
 \mathcal S_\vartheta H(\eta,\chi)
 =\int_0^{\re^{-\ri\vartheta}\infty}
 \re^{-\eta s}\widehat H(s,\chi)\,\dd s.
 \label{eq:short-directional-borel-sum}
\end{equation}
For \eqref{eq:short-directional-borel-sum} to be well defined, it is necessary that $\widehat H$ does not have singularities along the direction of integration. If there are singularities we define lateral Borel resummation as 
\begin{equation}
 \mathcal S_{\vartheta\pm}H
 =\lim_{\epsilon\to0^+}\mathcal S_{\vartheta\pm\epsilon}H.
 \label{eq:short-lateral-borel-sums}
\end{equation}
Borel summability alone does not guarantee that the Borel sum captures all non-perturbative effects, particularly in the presence of complex instantons, see e.g.~\cite{Grassi:2014cla}. However, the Nekrasov--Shatashvili representation of thermal correlators
\cite{AminovGrassiHatsuda:2020,Dodelson:2022yvn,JiaRangamani:2024Blocks,BhattaMandal:2023Exact,BhattaEtAl:2023Hyperbolic,ArnaudoGrassiHao:2025} motivates the existence of a chamber where an exact Borel reconstruction should be possible \cite{Hollands:2013qza,Grassi:2019coc,Grassi:2021wpw}.

To understand the reconstruction question using exact WKB analysis, we need two independent ingredients.  Our WKB conventions are summarized in Sec.~\ref{app:short-WKB-conventions} of the Supplementary Material.

First, we need to relate the exact WKB solution with the Frobenius solutions at the boundary. The WKB solutions normalized at the AdS boundary, $\psi_{\pm,\bdy}$, have non-trivial mixing with the Frobenius solutions $\psi_{\src,\res}$.
 It can be shown that this mixing captures the perturbative OPE expansion of $H$, which is the key mechanism underlying its bulk computation \cite{WIP,AfkhamiJeddi:2025bh}. For our purpose, we further need an exact WKB upgrade of this property. Let $\Psi_{\pm,\bdy}$ be the Borel sum of the formal WKB solutions $\psi_{\pm,\bdy}$, which is known to be well-defined in a Stokes region adjacent to AdS boundary \cite{IwakiNakanishi:2016wkbSimplePoles}. It can be shown that the  exact WKB generalization of the mixing is \cite{WIP}
\begin{equation}
\begin{pmatrix}
\Psi_{+,\bdy}^{(\vartheta)}\\
\Psi_{-,\bdy}^{(\vartheta)}
\end{pmatrix}
=
\begin{pmatrix}
\ri \re^{-\ri\pi\nu}a_{\mathrm{src}}^{(\vartheta)}
&
i \re^{\ri\pi\nu}a_{\mathrm{res}}^{(\vartheta)}
\\
a_{\mathrm{src}}^{(\vartheta)}
&
a_{\mathrm{res}}^{(\vartheta)}
\end{pmatrix}
\begin{pmatrix}
\psi_{\mathrm{src}}\\
\psi_{\mathrm{res}}
\end{pmatrix}.\label{eq:wkb_frob}
\end{equation}
Here the mixing coefficients $a_{\src/\res}$ are themselves Borel summed asymptotic expansions in inverse powers of $\beta\omega$. Their precise relation to the OPE expansion is \cite{WIP}
\begin{equation}
\frac{\mathcal N_\Delta}{G_{\mathrm R}^{\vac}}
\frac{a_{\mathrm{res}}^{(\vartheta)}}
     {a_{\mathrm{src}}^{(\vartheta)}}
=
\exp\!\left[\mathcal S_\vartheta H\right].
\label{eq:WKB-mixing-OPE}
\end{equation}
This is a completely local property at the AdS boundary.

Second, we transport the boundary-normalized WKB solutions to the horizon using standard Stokes connection formulae, as detailed in Sec.~\ref{app:short-WKB-transport} of the Supplementary Material. Combining this transport with \eqref{eq:WKB-mixing-OPE} gives the factorized expression \eqref{eq:short-reconstruction-retarded-general}: the vacuum-normalized retarded correlator is the Borel-resummed thermal OPE multiplied by a factor determined by global Stokes data~\cite{JiaRangamaniSingularity}. 
From \eqref{eq:short-reconstruction-upper-triangular}, it follows that exact Borel reconstruction holds precisely when
\begin{equation}
\Psi_{-,\bdy}^{(\vartheta)}
\propto
\Psi_{-,\mathrm{hor}}^{(\vartheta)},
\label{eq:short-reconstruction-condition}
\end{equation}
where our choice of branch for $\sqrt{Q_0}$ is such that $\Psi_{-,\mathrm{hor}}^{(\vartheta)}\propto\psi_{\mathrm{in}}$.
A sufficient condition for \eqref{eq:short-reconstruction-condition} to hold is that the minus WKB solution $\Psi_-$ is preserved at every Stokes crossing. This holds if each Stokes curve that we cross at $z_{\mathrm  c}$, emanating from a Stokes source $a$, satisfies
\begin{equation}
\operatorname{Re}\!\left[
\re^{\ri\vartheta}\int_a^{z_{\mathrm  c}}\lambda_0
\right]>0.
\label{eq:short-minus-preserving-Stokes-crossing}
\end{equation}
Each factor is then upper triangular in the ordered basis $(\Psi_+,\Psi_-)$, and their product preserves the minus solution.

.

\begin{figure*}[tbp]
 \centering
 \includegraphics[width=\textwidth]{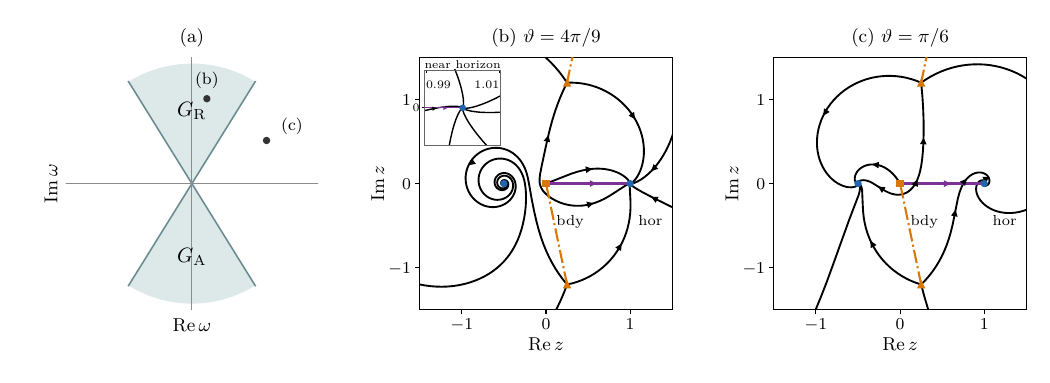}
 \caption{
 Retarded and advanced reconstruction wedges (a), and Stokes geometry inside (b) and outside (c) the retarded wedge, for $d=4$, $a_K=1$, $\chi=1/2$. Orange triangles mark the bulk turning points, the orange square marks the AdS boundary at $z=0$, and blue dots mark the double poles of $Q_0$ at $z=1$ (the physical horizon) and $z=-1/2$.
 The purple path denotes the WKB transport from boundary to horizon. The black lines are Stokes trajectories; their arrows point along increasing $\operatorname{Re}[\re^{\ri\vartheta}\int_a^z\lambda_0]$, where $a$ is the corresponding Stokes source. 
The orange dash-dotted curves are branch cuts. In (b), the crossed trajectories are oriented away from their sources: the positivity condition~\eqref{eq:short-minus-preserving-Stokes-crossing} holds at every crossing, preserving $\Psi_-$. In (c), the first crossed trajectory is oriented toward its source, so the same real part is negative and $\Psi_-$ mixes with $\Psi_+$. The inset in (b) resolves the first near-horizon crossing; infinitely many further crossings accumulate at the horizon, all preserving the minus solution.
 }
 \label{fig:reconstruction-stokes}
\end{figure*}
In Regime I, by inspection of the Stokes geometry (see Fig.~\ref{fig:reconstruction-stokes}), we find that this sufficient condition is satisfied in an upper wedge of the complex-frequency plane. 
The boundary of this chamber is associated with the $j=0$ Borel singularity in \eqref{eq:short-brane-regime-i-borel-singularities} and \eqref{eq:short-bh-turning-actions-general}:
\begin{equation}
\vartheta_*^{(0)}
=
-\arg s_{*,+}^{(0)}(\bdy,t_0),
\qquad
0<\vartheta_{*}^{(0)}<\frac{\pi}{2}.
\end{equation}
The exact reconstruction statement is therefore
\begin{equation}
  \boxed{
\begin{gathered}
\frac{G_{\mathrm R}}{G_{\mathrm R}^{\vac}}
=
\exp\!\left[\mathcal S_\vartheta H\right],
\\
0\leq\chi<\chi^{\min}_{\mathrm c},
\qquad
\vartheta_*^{(0)}
<
\vartheta
<
\pi-\vartheta_*^{(0)}.
\end{gathered}
  }
  \label{eq:short-reconstruction}
\end{equation}
In the planar limit, this is the timelike momentum regime, with reconstruction wedge
\begin{equation}
\frac{\pi}{2}-\frac{\pi}{d}
<
\vartheta
<
\frac{\pi}{2}+\frac{\pi}{d}.
\end{equation}
We note that the $d=2$ limit of this, where the reconstruction wedge is the upper half-plane, can indeed be verified directly from the \CFT{2} thermal two-point function fixed by conformal symmetry; all Borel singularities are real in this case. 
Numerical cross-checks of~\eqref{eq:short-reconstruction} are presented in Sec.~\ref{app:short-resummation} of the Supplementary Material.

The advanced correlator
is similarly reconstructed in the reflected lower wedge
\begin{equation}
\begin{gathered}
\frac{G_{\mathrm A}}{G_{\mathrm A}^{\vac}}
=
\exp\!\left[\mathcal S_\vartheta H\right],
\\
0\leq\chi<\chi_{\mathrm c}^{\min},
\qquad
-\pi+\vartheta_*^{(0)}
<
\vartheta
<
-\vartheta_*^{(0)}.
\end{gathered}
\end{equation}

The Borel sum of the thermal OPE exhibits Stokes jumps due to the Borel singularities. For the Borel singularities associated with boundary-to-turning-point saddle trajectories, we derive a jump formula analogous to the DDP jump formula for quantum WKB periods \cite{DelabaerePham:1999resurgent,DelabaereDillingerPham:1993,DelabaereDillingerPham:1997,KoikeTakei:2011whittaker}. For this purpose, it is useful to lift the saddle trajectory $l_*(\bdy, t_j)$ to a saddle period $\gamma_*$ on the WKB curve, so that
\begin{equation}
  s_* = \oint_{\gamma_*} \lambda_0.
\end{equation}
While the locations of the Borel singularities are determined by classical periods, their associated Stokes discontinuity involves the full quantum period 
\begin{equation}
  V_* = \oint_{\gamma_*} \lambda_{\mathrm  WKB}.
\end{equation}  
\label{eq:quantum-period}
Our jump formula states \cite{WIP}
\begin{equation}
  \boxed{
\exp\!\left[
\mathcal S_{\vartheta_*^-}H
-
\mathcal S_{\vartheta_*^+}H
\right]
=
\frac{1+\re^{-i\pi\nu-\mathcal V_*}}
     {1+\re^{\ri\pi\nu-\mathcal V_*}}.
  }
  \label{eq:short-jump-formula}
\end{equation}
Here $\mathcal V_*$ denotes the Borel sum of $V_*$ in direction $\vartheta_*$, and the orientation of the saddle cycle is chosen such that $\re^{-V_*}$ is exponentially small. The formula holds for any boundary-to-turning-point saddle trajectory, as long as $\chi$ lies away from the discriminant locus of the classical WKB curve. For example, although the reconstruction formula~\eqref{eq:short-reconstruction} holds also for $\chi=0$, the jump formula \eqref{eq:short-jump-formula}  at $\chi=0$ requires separate treatment \cite{WIP}.

\section{Discussion and outlook}

Our results show how the high-energy expansion of thermal two-point functions can encode both bulk geometrical quantities and, in a definite complex-frequency chamber, the full thermal response.

A natural question is how this resurgent momentum-space description relates to the position-space analysis. In position space, singularities of the stress-tensor OPE contribution have been associated with bulk geodesics \cite{Parisini:2023nbd, Ceplak:2024bja,Araya:2026shz,Jia:2026pmv}. In our momentum-space analysis, Borel singularities are determined by classical WKB actions along Stokes saddle trajectories connecting the AdS boundary to bulk turning points or, in certain regimes, to the black hole singularity. A natural next step is to understand how these Borel singularities are related to the  singularities in position space.

This analogy also raises the question of how the double-twist contributions required in position space are encoded in the resurgent momentum-space description. In position space, the stress-tensor contribution defines a series with a finite radius of convergence. Double-twist contributions cancel those singularities that are incompatible with thermal analyticity \cite{Ceplak:2024bja}. In momentum space, the limitations of the perturbative stress-tensor expansion instead manifest themselves in its asymptotic character: the large-frequency series alone does not uniquely determine the full correlator because it is asymptotic. This suggests that double-twist data encode the non-perturbative information associated with completing this asymptotic expansion, as also suggested in \cite{DodelsonIossaKarlssonZhiboedov:2024Product}. Understanding precisely how these data enter the resummation and its non-perturbative completion remains an open question.

It would be interesting to derive our reconstruction statement \eqref{eq:short-reconstruction} and jump formula \eqref{eq:short-jump-formula} directly from CFT without relying on bulk WKB arguments. Recent approaches relating thermal OPE data to the analytic structure and quasinormal spectrum of thermal correlators \cite{BarratEtAl:2026MomentumBootstrap,ArnaudoIossaKarlssonWithers:2026OPEQNM} may provide a starting point for addressing these questions. A field-theoretic derivation could clarify which aspects of this reconstruction mechanism extend beyond theories admitting a classical gravity description.

\section*{acknowledgments}

We  thank R.~Karlsson, M.~Kulaxizi, A.~Parnachev, M.~Rangamani and  B.~Withers for useful discussions and correspondence.
AG acknowledges support by the SNSF Grant No.~218510 and the NCCR SwissMAP. H.F.J. acknowledges support from the Tsinghua Shuimu Scholar program and hospitality from CERN where part of this work was completed. We used ChatGPT Pro to assist with calculations and manuscript editing. The authors take full responsibility for the content of this manuscript.

\FloatBarrier
\bibliographystyle{apsrev4-2-prl}
\bibliography{biblio}

\clearpage
\onecolumngrid
\begin{center}
\large\bfseries Supplementary Material
\end{center}
\raggedbottom
\setcounter{section}{0}
\setcounter{equation}{0}
\setcounter{figure}{0}
\setcounter{table}{0}
\renewcommand{\theequation}{S\arabic{equation}}
\renewcommand{\thefigure}{S\arabic{figure}}
\renewcommand{\thetable}{S\arabic{table}}

\section{Spherical black holes: Borel singularities}
\label{app:short-spherical-examples}
In this section, we discuss the Borel singularities for the spherical black hole in the regimes defined in \eqref{eq:regimes_borel}.
Here $f(z)=1+a_Kz-(1+a_K)z^{d/2}$.
\paragraph{Regime I.}
For $0<\chi<\chi^{\min}_{\mathrm  c}$, the primitive principal-sheet singularities
come from boundary-to-turning-point saddles and are given by
\begin{equation}
 \begin{aligned}
s_{*,\pm}^{(j)}(\bdy,t_j)
 &=\pm\int_0^{t_j}
 \frac{\sqrt{\chi^2f(z)-1}}{\sqrt z\,f(z)}\,\dd z.
 \end{aligned}
 \label{eq:short-bh-turning-actions-general}
\end{equation}
Here $j=0,\ldots,d/2-1$. 
\paragraph{Regime II.} The largest positive real turning-point collision value is
\begin{equation}
\chi_{\mathrm c}^{\max}=
\begin{cases}
\left[2-\left(\chi_{\mathrm c}^{\min}\right)^{-2}\right]^{-1/2},
&
d\equiv2\pmod4,
\quad
\chi_{\mathrm c}^{\min}>1/\sqrt2,
\\[2mm]
1,
&\text{otherwise}.
\end{cases}\label{eq:chi_max}
\end{equation}
For $\chi>\chi^{\max}_{\mathrm  c}$, all the singularities
$s_{*,\pm}^{(j)}(\bdy,t_j)$ are still present.
In addition, there are
boundary-to-singularity saddles with periods
\begin{equation}
 \begin{aligned}
s_{*,\pm}^{(j)}(\bdy,\sing)&=\pm \re^{\ri\alpha_j/2}\int_0^\infty\dd r\\
 &\times
 \frac{\sqrt{\chi^2[1+a_K\re^{\ri\alpha_j}r+(1+a_K)r^{d/2}]-1}}
 {\sqrt r\,[1+a_K\re^{\ri\alpha_j}r+(1+a_K)r^{d/2}]},
 \end{aligned}
 \label{eq:short-bh-infinity-ray-actions}
\end{equation}
where
$\alpha_j=2\pi(2j+1)/d\pmod{2\pi} \in (-\pi,\pi]$
with $j=0,\ldots,d/2-1$. 

Examples are given in Fig.~\ref{fig:borel-saddle-correspondence}.

\section{WKB solutions: definitions}
\label{app:short-WKB-conventions}
We select the sheet by
\begin{equation}
 \sqrt{Q_0(z)}\sim-\frac{\ri\sqrt{1-\chi^2}}{2\sqrt z}
 \quad(z\to0),
 \qquad
 \eta\sqrt{Q_0(z)}\sim\frac{\ri\beta\omega}{4\pi(z-1)}
 \quad(z\to1).
 \label{eq:short-reconstruction-WKB-sheet}
\end{equation}
The subleading potential is given by
\begin{equation}
\begin{aligned}
Q_2(z)
={}&
\frac{\nu^2-d^2/4}{4z^2f(z)}
+\frac{d^2-4}{16z^2}
-\frac{(d-2)a_K}{4zf(z)}
-\frac{f'(z)^2}{4f(z)^2},
\end{aligned}
\label{eq:subleading}
\end{equation}
where $f$ is given in \eqref{eq:metric}.

\subsection{Formal WKB solutions and their resummations}
Let $q$ be a turning point or a simple pole of $Q_0$.
The standard formal WKB solutions to \eqref{eq:short-radial-WKB} are given by
\begin{equation}
 \psi_{\pm,q}(z,\eta)
 =\frac{1}{\sqrt{S_{\mathrm  odd}(z,\eta)}}
  \exp\!\left(\pm\int_{q}^{z}S_{\mathrm  odd}(\zeta,\eta)\,\dd\zeta\right).
 \label{eq:short-formal-WKB-solutions}
\end{equation}
where \begin{equation}
 S_{\mathrm  odd}=\frac{S^{(+)}-S^{(-)}}2,
 \qquad
 S_{\mathrm  even}=\frac{S^{(+)}+S^{(-)}}2.
 \label{eq:short-odd-even-Riccati}
\end{equation}
and $S^{\pm}$ solve the Riccati equation
\be \partial_zS+S^2=\eta^2Q_0+Q_2.\ee
This has two formal solutions
\begin{equation}
 S^{(\pm)}(z,\eta)=\sum_{n=-1}^{\infty}\eta^{-n}S_n^{(\pm)}(z),
 \qquad
 S_{-1}^{(\pm)}=\pm\sqrt{Q_0},
 \qquad
 S_0^{(\pm)}=-\frac14\partial_z\log Q_0,
 \label{eq:short-formal-Riccati-branches}
\end{equation}
where the higher coefficients  $S_n^{\pm}$ are computed recursively. 
One can show that, term by term the integrals in \eqref{eq:short-formal-WKB-solutions} are well defined, see eg
\cite[Eqs.~(2.26)--(2.27)]{IwakiNakanishi:2014wkb} for simple turning points and \cite[Sec.~2.2]{IwakiNakanishi:2016wkbSimplePoles} for simple poles.

It is useful to introduce the all-order WKB differential and its regular part as
\begin{equation}
 \lambda_{\mathrm  WKB}=S_{\mathrm  odd}(z,\eta)\,\dd z,
 \qquad
 \lambda_{\mathrm  WKB}^{\mathrm  reg}
 =\lambda_{\mathrm  WKB}-\eta\lambda_0.
 \label{eq:short-WKB-differentials}
\end{equation}

For normalization points $a,b$ that are simple turning points or simple poles, changing the base point in \eqref{eq:short-formal-WKB-solutions} gives
\begin{equation}
 \psi_{\pm,a}(z,\eta)=Y_{ab}(\eta)^{\pm1}\psi_{\pm,b}(z,\eta),
 \qquad
 Y_{ab}(\eta)=\exp\!\left(\int_{\ell_{ab}}
 \lambda_{\mathrm  WKB}\right),
 \label{eq:short-formal-normalization-change}
\end{equation}
where $\ell_{ab}$ runs from $a$ to $b$. For such endpoints, one often uses the closed
Voros symbol $X_{ab}=\exp(\oint_{\gamma_{ab}}\lambda_{\mathrm  WKB})=Y_{ab}^2$.
Here $\gamma_{ab}$ is the closed cycle on the WKB surface obtained by
following $\ell_{ab}$ from $a$ to $b$ on one sheet and returning on the
other.

The formal solutions \eqref{eq:short-formal-WKB-solutions} describe
asymptotic expansions of exact solutions as $\eta\to\infty$ in suitable
sectors, see \cite{Nikolaev:2023ExactWKB} for a
rigorous study. To obtain exact solutions, we Borel resum them.
It is convenient to define
\begin{equation}
A_q(z)=\int_q^z\lambda_0~,
\end{equation}
and write
\begin{equation}
 \begin{aligned}
 \psi_{\pm,q}(z,\eta)
 &=\eta^{-1/2}Q_0(z)^{-1/4}\re^{\pm\eta A_q(z)}
   \bigl(1+\psi_{\pm,q}^{\mathrm  red}(z,\eta)\bigr),\\
 \psi_{\pm,q}^{\mathrm  red}(z,\eta)&=O(\eta^{-1}).
 \end{aligned}
 \label{eq:short-reduced-formal-WKB}
\end{equation}
For $z$ in a Stokes region where the reduced series is Borel summable in
direction $\vartheta$, and for $\eta$ such that the Laplace integral
converges, we define the analytic WKB solutions by
\begin{equation}
 \Psi_{\pm,q}^{\vartheta}(z,\eta)
 =\eta^{-1/2}Q_0(z)^{-1/4}\re^{\pm\eta A_q(z)}
   \bigl(1+\mathcal S_\vartheta\psi_{\pm,q}^{\mathrm  red}(z,\eta)\bigr)~.
 \label{eq:short-Borel-summed-WKB-solutions}
\end{equation}
where $\mathcal S_\vartheta$ is defined by
\eqref{eq:short-directional-borel-sum}, applied to
$\psi_{\pm,q}^{\mathrm  red}$.
If the summation direction is singular, we use the lateral sums as in
\eqref{eq:short-lateral-borel-sums}.

If $q$ is a double pole of $Q_0$ with $Q_2(z)=-1/[4(z-q)^2]+O((z-q)^{-1})$, as at the horizon $q=1$, the pole-normalized formal solutions are instead \cite[Eq.~(2.28)]{IwakiNakanishi:2014wkb}
\begin{equation}
 \psi_{\pm,q}(z,\eta)
 =\frac{1}{\sqrt{S_{\mathrm  odd}(z,\eta)}}
 \exp\!\left\{\pm\left[
 \eta\int_0^z\lambda_0
 +\int_q^z\lambda_{\mathrm  WKB}^{\mathrm  reg}
 \right]\right\}.
 \label{eq:short-double-pole-normalization}
\end{equation}
As before we can Borel resum this expression, and this defines $\Psi_{\pm,q}^{(\vartheta)}$.

With the branch convention \eqref{eq:short-reconstruction-WKB-sheet}, the Borel-summed WKB solutions behave near the horizon as
\begin{equation}
\Psi_{\pm,\hor}^{(\vartheta)}(z,\eta)\propto
 (1-z)^{1/2\pm\ri\beta\omega/(4\pi)},
 \qquad z\to1.
 \label{eq:short-horizon-WKB-asymptotics}
\end{equation}
Comparing with the ingoing and outgoing Frobenius solutions defined in \eqref{eq:short-horizon-Frobenius}, we obtain
\begin{equation}
\Psi_{-,\hor}^{(\vartheta)}=b_{\mathrm  in}\psi_{\mathrm  in},
 \quad
\Psi_{+,\hor}^{(\vartheta)}=b_{\mathrm  out}\psi_{\mathrm  out},
 \label{eq:short-reconstruction-horizon-matching}
\end{equation}
where $b_{\mathrm  in/out}$ are normalization factors independent of $z$.

\section{Reconstruction wedge: technical details}
\label{app:short-WKB-transport}

A technical ingredient in the derivation of the reconstruction statement \eqref{eq:short-reconstruction} is the WKB transport between the boundary and the horizon, which we review here. 

The main idea is that we can relate the WKB solution at the boundary and at the horizon by appropriately taking into account all the Stokes jumps connecting them.
Specifically we have  the following transport relation \cite{Voros:1983quartic,AokiKawaiTakei:1991wkb,Koike:2000wkb,IwakiNakanishi:2014wkb,IwakiNakanishi:2016wkbSimplePoles} 
\begin{equation}
\begin{pmatrix}\Psi_{+,\bdy}^{(\vartheta)}\\\Psi_{-,\bdy}^{(\vartheta)}\end{pmatrix}
 =\mathbb K^{(\vartheta)}
 D\!\left(\mathcal E_{\bdy,\hor}^{(\vartheta)}\right)
\begin{pmatrix}\Psi_{+,\hor}^{(\vartheta)}\\\Psi_{-,\hor}^{(\vartheta)}\end{pmatrix},
 \qquad
D(e)=\begin{pmatrix}e&0\\0&\re^{-1}\end{pmatrix}.
 \label{eq:short-reconstruction-WKB-transport}
\end{equation}
Here $\mathbb K^{(\vartheta)}$ is the ordered product of crossed Stokes factors. The factor $\mathcal E_{\bdy,\hor}^{(\vartheta)}$ compares the boundary and horizon normalizations, parallel to \eqref{eq:short-formal-normalization-change} and reads  \cite[Sec.~2.2]{IwakiNakanishi:2016wkbSimplePoles}\begin{equation}
 \begin{aligned}
 V_{\bdy,\hor}^{\mathrm  reg}(\eta,\chi)
 &=\int_0^1\lambda_{\mathrm  WKB}^{\mathrm  reg},\\
 \mathcal E_{\bdy,\hor}^{(\vartheta)}
 &=\exp\!\left[\mathcal S_\vartheta V_{\bdy,\hor}^{\mathrm  reg}\right].
 \end{aligned}
 \label{eq:short-open-normalization-Borel}
\end{equation}
For transport across Stokes lines emanating from simple turning points or simple poles, the transport matrix $\mathbb K^{(\vartheta)}$ takes the following form
\cite{Voros:1983quartic,AokiKawaiTakei:1991wkb,Koike:2000wkb,IwakiNakanishi:2014wkb,IwakiNakanishi:2016wkbSimplePoles}
\begin{equation}
 \mathbb K^{(\vartheta)}=\prod_i K_i(\epsilon_i s_i),
 \qquad
 U(s)=\begin{pmatrix}1&s\\0&1\end{pmatrix},
 \qquad
 L(s)=\begin{pmatrix}1&0\\s&1\end{pmatrix},
 \quad K_i\in\{U,L\}~.
 \label{eq:short-reconstruction-Stokes-factors}
\end{equation}
Here $\epsilon_i=\pm1$ records the crossing orientation, and $s_i$ includes the Stokes multiplier and the Voros factors for the chosen normalization. A Stokes curve from $a_i$ crossed at $z_{{\mathrm  c},i}$ contributes $U$ if $\operatorname{Re}[\re^{\ri\vartheta}\int_{a_i}^{z_{{\mathrm  c},i}}\lambda_0]>0$, and $L$ if this real part is negative; these preserve $\Psi_-$ and $\Psi_+$, respectively \cite[Eqs.~(2.19), (2.21)]{IwakiNakanishi:2016wkbSimplePoles}. 

Combining \eqref{eq:short-reconstruction-WKB-transport} and \eqref{eq:short-reconstruction-horizon-matching} with the boundary relations \eqref{eq:wkb_frob}--\eqref{eq:WKB-mixing-OPE} and the holographic prescription \eqref{short-GR-holo-prescription}, we obtain
\begin{equation}
 \frac{G_{\mathrm  R}}{G_{\mathrm  R}^{\vac}}
 =\exp\!\left[\mathcal S_\vartheta H\right]
 \frac{(\mathbb K^{(\vartheta)})_{11}
       -\ri\re^{\pi\ri\nu}(\mathbb K^{(\vartheta)})_{21}}
      {(\mathbb K^{(\vartheta)})_{11}
       -\ri\re^{-\pi\ri\nu}(\mathbb K^{(\vartheta)})_{21}}.
 \label{eq:short-reconstruction-retarded-general}
\end{equation}
Hence for $\nu\notin\mathbb Z$, the Borel sum reconstructs the full retarded correlator if and only if $\mathbb K^{(\vartheta)}$ is upper triangular,
\begin{equation}
 (\mathbb K^{(\vartheta)})_{21}=0.
 \label{eq:short-reconstruction-ingoing-criterion}
\end{equation}
A sufficient condition is that every crossed Stokes factor is upper triangular. In this case
\begin{equation}
 \mathbb K^{(\vartheta)}=\prod_jU(\epsilon_js_j)
 =U\!\left(\sum_j\epsilon_js_j\right).
 \label{eq:short-reconstruction-upper-triangular}
\end{equation}
With the branch choice \eqref{eq:short-reconstruction-WKB-sheet}, this is ensured by the positivity condition \eqref{eq:short-minus-preserving-Stokes-crossing}, as illustrated in Fig.~\ref{fig:reconstruction-stokes}.

\section{Borel resummation}
\label{app:short-resummation}
We compare Borel resummation with the independent numerical solutions of
the radial connection problem for neutral black branes and black holes in five and seven
bulk dimensions. The parameters are
\begin{equation}
 \begin{aligned}
 & \nu=\frac{13}{10},
 ~ &\chi\in\left\{\frac35,0\right\},
 ~ ~&\vartheta_{d=4}=\frac{3\pi}{8},
  ~&\vartheta_{d=6}=\frac{5\pi}{12}.
 \end{aligned}
 \label{eq:short-planar-numerical-parameters}
\end{equation}
For the black brane we take $\eta=2\re^{\ri \vartheta_d}$ while for the black hole we set $\eta=5\re^{\ri \vartheta_d}$ and $a_K=1$.

We also use
\begin{equation}
     G_{\mathrm  R}^{\vac}(\omega,\lambda)
 =-\mathcal N_\Delta
 \left({r_+}\right)^{2\nu}
 \re^{-\pi\ri\nu}\frac{\Gamma(1-\nu)}{\Gamma(1+\nu)}
 \left[\frac{\eta^2}{4}(1-\chi^2)\right]^\nu\qquad \nu\notin\mathbb Z.
 \label{eq:short-vacuum-normalization}
\end{equation}
The  branch is fixed by taking $\sqrt{1-\chi^2}>0$ for
$0\leq\chi<1$ and $\eta>0$, and then analytically continuing
to the other regions. 
The numerical method is analogous to the one in \cite{AfkhamiJeddi:2025bh}.
The results are given in Tables~\ref{tab:reconstruction-planar-d4}--\ref{tab:short-spherical-resummation-bulk-seven}.

\begin{table}[H]
 \centering
 \small
 \renewcommand{\arraystretch}{1.15}
 \setlength{\tabcolsep}{4pt}
 \begin{tabular}{@{}r@{\hspace{0.02\textwidth}}c@{\hspace{0.02\textwidth}}c@{}}
 \hline
 \multicolumn{3}{c}{Borel resummation $\log[G_R^\vartheta/G_{\mathrm R}^{\vac}]$} \\
 \makebox[0.06\textwidth][r]{\(N\)} & \makebox[0.45\textwidth][c]{\(\chi=3/5\)} & \makebox[0.45\textwidth][c]{\(\chi=0\)} \\
 \hline
 16 & \(-0.0400788879871570- \underline{0.110}7426642453942\ri\) & \(-\underline{0.00913}84816735057- \underline{0.038817}8294117939\ri\)\\
 64 & \(-\underline{0.03981}08566975457- \underline{0.11043}50705560881\ri\) & \(-\underline{0.0091369}414894854- \underline{0.03881744}29720129\ri\)\\
 128 & \(-\underline{0.039811}0298315385- \underline{0.110431}9641550767\ri\) & \(-\underline{0.0091369292}913222- \underline{0.038817445}9700062\ri\)\\
 512 & \(-\underline{0.03981115}82042325- \underline{0.110431611}9023015\ri\) & \(-\underline{0.009136929244094}9- \underline{0.03881744586978}30\ri\)\\
 1024 & \(-\underline{0.03981115792}76521- \underline{0.11043161148}56752\ri\) & \(-\underline{0.0091369292440947}- \underline{0.0388174458697826}\ri\)\\
 \hline
 ODE & \(-0.0398111579280670- 0.1104316114847638\ri\) & \(-0.0091369292440947- 0.0388174458697826\ri\)\\
 \hline
 \end{tabular}
 \caption{Agreement between the Borel resummation of 
 $G_{\mathrm  R}^{(\vartheta)}$ and the ODE numerical value
for  black brane in \AdS{5}. We underline stable digits. 
 Here $N$ is the number of  non-zero coefficients in the perturbative series. For even
 $N$, the  Borel transform is approximated by a
 $[N/2-1\,/\,N/2]$ Pad\'e approximant before evaluating the
 Laplace integral \eqref{eq:short-directional-borel-sum}.}
 \label{tab:reconstruction-planar-d4}
\end{table}

\begin{table}[H]
 \centering
 \small
 \renewcommand{\arraystretch}{1.15}
 \setlength{\tabcolsep}{4pt}
 \begin{tabular}{@{}r@{\hspace{0.02\textwidth}}c@{\hspace{0.02\textwidth}}c@{}}
 \hline
 \multicolumn{3}{c}{Borel resummation $\log[G_R^\vartheta/G_{\mathrm R}^{\vac}]$} \\
 \makebox[0.06\textwidth][r]{\(N\)} & \makebox[0.45\textwidth][c]{\(\chi=3/5\)} & \makebox[0.45\textwidth][c]{\(\chi=0\)} \\
 \hline
 16 & \(-\underline{0.05}8472181933- \underline{0.08}0712097501\ri\) & \(-\underline{0.015}4024033037941- \underline{0.0296}352266271579\ri\)\\
 64 & \(-\underline{0.056}437419310- \underline{0.08380}3522321\ri\) & \(-\underline{0.01523}56482297377- \underline{0.0296}918742878144\ri\)\\
 128 & \(-\underline{0.0560}58087931- \underline{0.0838}43758593\ri\) & \(-\underline{0.015232}7549903440- \underline{0.029688}5315603043\ri\)\\
 1024 & \(-\underline{0.056008}925211- \underline{0.083808}154171\ri\) & \(-\underline{0.0152326533}630992- \underline{0.029688485}4651270\ri\)\\
 4096 & \(-\underline{0.056008139}684- \underline{0.083808307}889\ri\) & \(-\underline{0.01523265339}15318- \underline{0.029688485390838}9\ri\)\\
 \hline
 ODE & \(-0.056008139842- 0.083808307928\ri\) & \(-0.0152326533916865- 0.0296884853908382\ri\)\\
 \hline
 \end{tabular}
 \caption{Agreement between the Borel resummation of 
 $G_{\mathrm  R}^{(\vartheta)}$ and the ODE numerical value
for  black brane in \AdS{7}.
 The meaning of $N$ is  as in Table \ref{tab:reconstruction-planar-d4}.}
 \label{tab:reconstruction-planar-d6}
\end{table}

\begin{table}[H]
 \centering
 \small
 \renewcommand{\arraystretch}{1.15}
 \setlength{\tabcolsep}{4pt}
 \begin{tabular}{@{}r@{\hspace{0.02\textwidth}}c@{\hspace{0.02\textwidth}}c@{}}
 \hline
 \multicolumn{3}{c}{Borel resummation \(\log[G_{\mathrm  R}/G_{\mathrm R}^{\vac}]\)} \\
 \makebox[0.06\textwidth][r]{\(N\)} & \makebox[0.45\textwidth][c]{\(\chi=3/5\)} & \makebox[0.45\textwidth][c]{\(\chi=0\)} \\
 \hline
 16 & \(\underline{0.083}4772515192937 + \underline{0.066}5809709587563\ri\) & \(\underline{0.04496}60186505208 + \underline{0.04026}78800565773\ri\)\\
 64 & \(\underline{0.083459}5009445817 + \underline{0.0666624}226130518\ri\) & \(\underline{0.0449660359}417116 + \underline{0.0402678569}548463\ri\)\\
 128 & \(\underline{0.083459}5008277428 + \underline{0.0666624}224707926\ri\) & \(\underline{0.0449660359421}614 + \underline{0.0402678569541}581\ri\)\\
 512 & \(\underline{0.083459500759}5689 + \underline{0.06666242236}56172\ri\) & \(\underline{0.044966035942161}4 + \underline{0.04026785695415}82\ri\)\\
 1024 & \(\underline{0.08345950075973}63 + \underline{0.06666242236567}55\ri\) & \(\underline{0.044966035942161}4 + \underline{0.04026785695415}82\ri\)\\
 \hline
 ODE & \(0.0834595007597359 + 0.0666624223656753\ri\) & \(0.0449660359421612 + 0.0402678569541580\ri\)\\
 \hline
 \end{tabular}
 \caption{
 Agreement between the Borel resummation
 of $G_{\mathrm  R}^{(\vartheta)}$ and the ODE numerical value
for  black hole in \AdS{5}. 
 The meaning of $N$ is  as in Table \ref{tab:reconstruction-planar-d4}.}

 \label{tab:short-spherical-resummation-bulk-five}
\end{table}

\begin{table}[H]
 \centering
 \small
 \renewcommand{\arraystretch}{1.15}
 \setlength{\tabcolsep}{4pt}
 \begin{tabular}{@{}r@{\hspace{0.02\textwidth}}c@{\hspace{0.02\textwidth}}c@{}}
 \hline
 \multicolumn{3}{c}{Borel resummation \(\log[G_{\mathrm  R}/G_{\mathrm R}^{\vac}]\)} \\
 \makebox[0.06\textwidth][r]{\(N\)} & \makebox[0.45\textwidth][c]{\(\chi=3/5\)} & \makebox[0.45\textwidth][c]{\(\chi=0\)} \\
 \hline
 16 & \(\underline{0.28}35473718583355 + \underline{0.13}68736516948159\ri\) & \(\underline{0.179}3508802363207 + \underline{0.09}29220190779876\ri\)\\
 64 & \(\underline{0.283}6748906682225 + \underline{0.1367}482485459195\ri\) & \(\underline{0.1793576}216664827 + \underline{0.0929207}177415493\ri\)\\
 128 & \(\underline{0.2836}631540249779 + \underline{0.1367}366011063481\ri\) & \(\underline{0.17935762}20469900 + \underline{0.09292071}62584019\ri\)\\
 512 & \(\underline{0.2836631}711024953 + \underline{0.1367366}485422213\ri\) & \(\underline{0.179357622084}4017 + \underline{0.09292071618}89814\ri\)\\
 1024 & \(\underline{0.28366317}20417869 + \underline{0.13673665}22693202\ri\) & \(\underline{0.17935762208452}78 + \underline{0.09292071618907}44\ri\)\\
 \hline
 ODE & \(0.2836631719699484 + 0.1367366523530372\ri\) & \(0.1793576220845269 + 0.0929207161890741\ri\)\\
 \hline
 \end{tabular}
 \caption{Agreement between the Borel resummation of 
 $G_{\mathrm  R}^{(\vartheta)}$ and the ODE numerical value
for  black hole in \AdS{7}. 
 The meaning of $N$ is  as in Table \ref{tab:reconstruction-planar-d4}.}

 \label{tab:short-spherical-resummation-bulk-seven}
\end{table}

\end{document}